\documentclass[11pt]{article}
\usepackage{amsfonts}
\usepackage{amssymb,latexsym}
\usepackage{amsmath,amscd}
\usepackage{theorem}
\usepackage[all]{xy}

\newcommand{\stk}[1]{\stackrel{#1}{\longrightarrow}}

\newcommand{\al}{\alpha}

\newcommand{\dwn}[1]{{\scriptstyle #1}\downarrow}

\def\RR{{\rm I}\!{\rm R}}

\def\NN{{\rm I}\!{\rm N}}

\def\Jo{J_{0}}
\def\Jotil{\widetilde{J}_{0}}

\def\Jtil{\widetilde{J}}

\def\J{\bf J}

\def\g{\bf g}

\def\gtil{\widetilde{\bf g}}

\def\Xtil{\widetilde{X}}
\def\rhotil{\widetilde{\rho}}
\def\H0{\widetilde{H}_{0}}

\def\fit{\widetilde{\phi}}

\def\gsom{\widehat{\g}}
\def\Jsom{\widehat{J}}
\def\d{\displaystyle}
\def\f{\forall}

\def\RR{{\rm I}\!{\rm R}}

\newtheorem{lemma} {Lemma} [section]
\newtheorem{proposition} [lemma] {Proposition}

\newtheorem{corollary} [lemma] {Corollary}
\newtheorem{definition}[lemma] {Definition}

\theorembodyfont{\rm}
\newtheorem{example}[lemma] {Example}
\newtheorem{remark}[lemma]{Remark}

\input{tcilatex}

\begin{document}

\title{\textbf{On Quantum Momentum Maps associated to non Ad$^*$-equivariant Classical Momentum Maps }}
%\author{Mar\'\i a Eugenia Garcia\thanks{Corresponding author. E-mail: \texttt{maru@mate.unlp.edu.ar}
%\newline \mbox{~~~~~} Tel.:+54-221-4229850 int 118 Fax:+54-221-4245875.
% and} \\
%\author{Marcela Zuccalli\thanks{Corresponding author. E-mail: \texttt{maru@mate.unlp.edu.ar}
%\newline \mbox{~~~~~} Tel.:+54-221-4229850 int 118 Fax:+54-221-4245875.
%} \\
%EndAName
%Departamento de Matem\'{a}tica \\
%Universidad de La Plata\\
%Calle 50 esq 115 (1900)\\
%La Plata, Argentina}
%\maketitle

\author{M.E. Garcia * \& M. Zuccalli*
\and * Departamento de Matem\'{a}tica, Universidad Nacional de La
Plata, \and Calle 50 esq. 115 (1900) La Plata, Argentina}
\maketitle

\begin{abstract}

In an interesting work M.F. M\"{u}ller-Bahns and N. Neumaier
({\newblock{\em "Some remarks on ${\g}$-invariant Fedosov star
products and quantum momentum mappings"}. Journal of Geometry and
Physics 50 (2004), 257-272.}) analyze the existence of a quantum
momentum map based on the existence of a classical momentum map
providing an answer to the proposal given by P. Xu in ({\newblock
{\em "Fedosov $*$-products and quantum momentum maps"}. Commun.
Math. Phys (1998) 167-197}). In both papers only equivariant
classical momentum maps are considered.

In these notes, we extend M\"{u}ller-Bahns and Neumaier analysis
to the case of a non equivariant momentum map. In addition, we
propose the notion of an anomalous quantum momentum map as an
alternative to recover a non equivariant momentum map at the
classical level by considering central extensions of the Lie
algebra associated with non equivariance.

\end{abstract}

MSC2000 Subject Classification Numbers:37C80; 37J15; 53D20; 70G65


\tableofcontents

\section{Introduction}

\ \\

In the last years, numerous papers have considered the relation
between classical and quantum symmetries in mechanical systems.
The fundamental role that the momentum map plays in the analysis
of classical mechanical systems with symmetries is well known. On
the other hand, deformation quantization provides a mathematical
framework for the problem of quantizing a classical mechanical
system. In this context, the quantum momentum map plays the role
of a quantum analogous to the classical momentum map.

In an interesting work \cite{Xu}, Xu has proved that a quantum
momentum map always recovers an $Ad^*$-equivariant classical
momentum map. He also raised the question whether the existence of
a classical momentum map guarantees the existence of a quantum
momentum map that recovers it at the classical limit.

Some years ago, M\"{u}ller-Bahns and Neumaier \cite{alemanes} have
given a negative answer to this question and have established the
necessary and sufficient conditions so that the existence of an
$Ad^*$-equivariant classical momentum map implies the existence of
a quantum momentum map associated.

In both works, they only considered $Ad^*$-equivariant classical
momentum maps. The aim of this paper is to generalize these kind
of ideas in order to include a non $Ad^*$-equivariant classical
momentum map that cannot be recovered at the classical limit of a
quantum momentum map.

Our approach is based on the introduction of the notion of an
anomalous quantum momentum map that recover a non
$Ad^*$-equivariant classical momentum map at the classical limit.
Although an anomalous quantum momentum map is not a Lie algebra
homomorphism, we establish conditions that guarantees it can be
considered as the restriction of one.

Then, we consider a classical mechanical system with a symmetry
given by a Lie group $G$ that admits a non $Ad^*$-equivariant
classical momentum map ${\J}_0$. In order to recover ${\J}_0$, we
define an anomalous quantum momentum map by two ways.

In the first place, we consider $\widetilde{\g}$, the central
extension of ${\bf g}$, the Lie algebra of $G$, defined by the
2-cocycle that measures the non $Ad^*$-equivariance of ${\J}_0$.
We define $\widetilde{\J}_0$ as the canonically extended classical
momentum map associated to the symmetry of the Lie algebra
$\widetilde{\g}$.

Following \cite{alemanes} we obtain  necessary and sufficient
conditions for the existence of a quantum momentum map
$\widetilde{\J}$ that recovers $\widetilde{\J}_0$. It is clear
that ${\J}_0$ can be recovered at the classical limit of the
restriction of $\widetilde{\J}$ to ${\bf g}$ that results an
anomalous quantum momentum map.

Furthermore, we establish necessary and sufficient conditions for
the existence of an anomalous quantum momentum map ${\bf J}$ whose
classical limit is ${\J}_0$. It is clear that ${\J}$ is not a Lie
algebra homomorphism. The 2-cocycle defined by ${\bf J}$ gives
rise to a central extension ${\bf \widehat{g}}$ of ${\bf g}$ by
$\RR[[\hbar]]$, the ${\bf g}$-module of the formal power series in
the parameter $\hbar$ with coefficients in $\RR$. We consider
${\bf \widehat{J}}$ the quantum momentum map that canonically
extends ${\bf J}$.

The present paper is organized as follows. In section 2 we recall
the definitions of an $Ad^*$-equivariant and a non
$Ad^*$-equivariant classical momentum map. Also we recall some
basic facts on Lie algebra extensions. In section 3 we give a very
short review of the deformation quantization and Fedosov's
construction. In section 4 we recall the definition of a quantum
momentum map and we collect some of the results proved in
\cite{alemanes}.

In section 5 we propose the notion of a anomalous quantum momentum
map. In section 6 we consider the case of non $Ad^*$-equivariant
classical momentum map and its canonically extended classical
momentum map that results equivariant. Also we analyze the quantum
momentum map associated to this equivariant momentum map. Finally,
in section 7, we study a canonical extended anomalous quantum
momentum map.

\bigskip

\section{Classical Momentum Maps}

\subsection{$Ad^*$-equivariant Classical Momentum Maps}

\hspace{0.4cm}The momentum map plays a fundamental role in the
analysis of classical mechanical systems with symmetries. In this
section, we recall its definition and some of its important
properties.
\ \\

Let us consider a symplectic manifold $(M,\omega)$ and
$C^{\infty}(M)$, the linear space of the differentiable functions
on $M$ with values in $\RR$. It is well known that $C^{\infty}(M)$
admits a canonical structure of Lie algebra associated to the form
$\omega$ given by the Poisson bracket defined as follows. Let
$X_{f}$ be the Hamiltonian vector field of $f$ given by the
condition $i_{X_{f}}\omega=d\, f$ with $i_X$ the contraction of
the 2-form $\omega$ by $X$ and $d$ the exterior differential
operator on $M$. If $f$ and $g \ \in C^{\infty}(M)$ then
$\{f,g\}=\omega(X_{f},X_{g})$.

The adjoint representation $ad_{\infty}$ of $C^{\infty}(M)$ on
itself is given by $(ad_{\infty})_f(\cdot) = \{f,\ \cdot\ \}$.

We consider a symplectic left action of a Lie group $G$ on $M$.
That is, there exists a differential application $\phi:G\times
M\rightarrow M$ such that $\phi^*_{g}\omega=\omega\ \ \forall \ g\
\in G$, where $\phi^*_{g}$ is the pull-back of the diffeomorphism
$\phi_{g}:M\rightarrow M$ given by $\phi_{g}(m)=\phi(g,m)$.

If $\g$ is the Lie algebra of the group $G$, $X_{\xi}$ denotes the
infinitesimal generator associated to the action $\phi$
corresponding to $\xi \in \g$ and $\g^*$ denotes the dual space of
$\g$.

An action of $G$ on $M$ canonically induces a representation
$\rho$ of the Lie algebra ${\bf g}$ on $C^{\infty}(M)$ defined as
${\displaystyle \rho(\xi)(f)=-L_{X_{\xi}}f}$, where $L$ denotes
the Lie derivative. Under such action, $C^{\infty}(M)$ becomes a
$\g$-module. In general, $\g$ acts on all differential forms on
$M$ in the same way; that is
$\rho(\xi)(\alpha)=-L_{X_{\xi}}\alpha$.

\begin{definition}
A differential function $J_0:M\rightarrow {\g}^*$ is a classical
momentum map for the action $\phi$ of $G$ on $M$ if
$$\langle J_0(m),\xi\rangle={\J}_0(\xi)(m) \ \ \forall\ m \in M \ \mbox{and}\  \xi\in{\g} $$
where ${\J}_0(\xi) \in C^{\infty}(M)$ satisfies that
$d{\J}_0(\xi)=i_{X_{\xi}}\omega \ \ \forall \ \xi \in \g.$

Thus, a momentum map can be considered as an application
${\J}_{0}:{\g}\rightarrow C^{\infty}(M)$ such that
$X_{{\J}_0(\xi)} = X_{\xi}\ \ \forall \ \xi \in \g$.
\end{definition}

\begin{remark}
If ${\J}_0$ is a momentum map for the action $\phi$ then

$$ \{{\J}_0(\xi),f\} = \omega(X_{\xi},X_{f}) = \rho(\xi)(f) =
(ad_{\infty})_{{\J}_0(\xi)}(f)\ \ \forall\ f\in C^{\infty}(M)$$
\end{remark}
\ \\

In order to consider the cohomology of $\g$ with coefficients in
$C^{\infty}(M)$ and other cohomologies derived from it, we recall
the definition of the cohomology of $\g$ with coefficients in a
$\g$-module $V$.

Given a linear space $V$ and $\rho$ a representation of $\g$ in
$V$, let us consider $C^{k}({\g},V)$ the space of alternate
$k$-multilineal maps $\alpha$ on $\g$ with values in $V$. The
Chevalley-Eilenberg coboundary operator associated
$\delta_{\rho}:C^{k}({\g},V)\rightarrow C^{k+1}({\g},V) \ \forall
\ k \in \ \NN$ is given by

\begin{eqnarray*}\delta_{\rho}(\alpha)(\xi_1,\xi_2,...,\xi_{k+1})
& = & \sum_{i=0}^{k+1} (-1)^{i+1}\rho(\xi_i) \left(
\alpha(\xi_1,\xi_2,...,\hat{\xi_i},...,\xi_{k+1}) \right) \\
& + & \sum_{i<j} \alpha \left( [\xi_i,\xi_j],
\xi_1,...,\hat{\xi_i},...,\hat{\xi_j},...,\xi_{k+1} \right)
\end{eqnarray*}

where the symbol $^{\wedge}$ means that the variable under it has
been deleted.

\ \\

An element $\alpha \in C^{k}({\g},V)$ is a $k$-cocycle if
$\delta_{\rho}(\alpha) = 0$ and it is a $k$-coboundarie if there
exists an element $\beta \in C^{k-1}({\g},V)$ such that
$\delta_{\rho}(\beta) = \alpha$.

If $Z^{k}_{\rho}({\g},V)$ is the space of the $k$-cocycles and
$B^{k}_{\rho}({\g},V)$ is the space of the $k$-coboundaries,

$$H^{k}_{\rho}({\g},V) = Z^{k}_{\rho}({\g},V)\ /\ B^{k}_{\rho}({\g},V)$$
is the $k$-group of the cohomology of $\g$ with coefficients in
$V$.
\ \\

\begin{remark}
Let us notice that ${\J}_{0} \in C^{1}({\g},C^{\infty}(M))$ where
$\rho$ is the representation of $\g$ on $C^{\infty}(M)$ as defined
above.
\end{remark}
\ \\

In order to define an $Ad^*$-equivariant momentum map, we recall
that the adjoint action $Ad$ of $G$ on $\g$ is given by $Ad_{g}
\xi = (d\, I_g)_{e} (\xi)$ with $g \in G$ and $\xi \in \g$, where
$I_g$ denotes the inner diffeomorphisms of $G$. Then, the
coadjoint action of $G$ on $\g^*$ is defined as
$(Ad^{*}_{g}\eta)\xi = \eta(Ad_{g}\xi) $ with $\xi \in \g $ and
$\eta \in \g^*$.

\begin{definition}

A classical momentum map $\Jo$ is $Ad^{*}$-equivariant if
$${\Jo}(\phi_{g}(m))=Ad^{*}_{g^{-1}}{\Jo}(m)\ \ \forall \ g \in G.$$

That is, $\Jo$ is $Ad^{*}$-equivariant if  the following diagram
is commutative

$$\begin{array}{ccc}
         M      & \stk{\phi_{g}}        & M\\
\dwn{\Jo}        &                       & \dwn{\Jo}\\
 {\g}^{*}&\stk{Ad^{*}_{g^{-1}}} & {\g}^{*}
\end{array}$$
\end{definition}

\begin{remark}
It is easy to see that if $\Jo$ is $Ad^*$-equivariant then
${\J}_0$ is a Lie algebra homomorphism. That is, ${\displaystyle
{\J}_0([\xi,\eta])=\{{\J}_0(\xi),{\J}_0(\eta)\}}$ for all
$\xi,\eta \in \g$.
\end{remark}

In the following subsections and sections, central extensions of
the Lie algebra of the symmetry group will need to be considered.
Now we shall provide the definition of the extension of a Lie
algebra and some of its important properties.
\ \\

If ${\g}$ is a Lie algebra and $V$ is an ${\g}$-module, an
extension $\overline{\g}$ of ${\g}$ by $V$ is a short exact
sequence

$$0 \rightarrow {\g} \rightarrow \overline{\g} \rightarrow V \rightarrow 0$$
where $\overline{\g}$ is identified with ${\g} \oplus V$.

It is well known that the extensions of a Lie algebra ${\g}$ by an
${\g}$-module $V$ are parametrized by the second cohomology group
$H^2({\g},V)$. Given $\Theta \in H^2({\g},V)$, the Lie commutator
on $\overline{\g}$ is defined as

$$[(a,v),(b,w)] = \left([a,b], a \cdot v - b \cdot w + \Theta (a,b)\right)\ \ \forall\ a,b\in{\g}\ \mbox{and}\ v,w\in V.$$

If the action of ${\g}$ on $V$ is trivial, the extension
$\overline{\g}$ is called central and its bracket is given by
$\displaystyle{[(a,v),(b,w)] = \left([a,b], \Theta (a,b)\right)}.$

\subsection{Non $Ad^*$-equivariant Classical Momentum
Maps}\label{subsection 2.2}

\hspace{0.4cm}In this subsection, we will consider the general
case of the classical mechanical systems with symmetry that admits
a non Ad$^*$-equivariant momentum map $J_0:M\rightarrow \g^*$. For
details consult \cite{A-M} and \cite{M-R}.

For $g \in G$ and $\xi \in \g$ we can consider the application
$\psi_{g,\xi}:M \rightarrow \RR$ given by

$$\psi_{g,\xi} (m) = {\J}_0(\xi)(\phi_g(m)) - {\J}_0(Ad_{g^{-1}}\xi)(m).$$

It is easy to see that this function is constant on $M$. Then, we
can define the map that measures the lack of $Ad^*$-equivariance
of the momentum ${\J}_0$.

\begin{definition}
Let us define $\sigma:G\rightarrow\g^*$ as the application  given
by $<\sigma(g),\xi> = \psi_{g,\xi}$.

That is, ${\displaystyle \sigma(g)=J_0(\phi_{g}(m))
-Ad^*_{g}(J_0(m)) \ \ \ \mbox{for some} \ \ m\in M}$.

\end{definition}

\begin{remark}

The map $\sigma$ satisfies $\sigma(g\, h) = \sigma(g) +
Ad^{*}\sigma(h)$ for all $g, h \in G$. Then, $\sigma$ is a
1-cocycle on $G$ with values in $\g^*$.

It is clear that $J_0$ is $Ad^*$-equivariant if and only if its
associated cocycle $\sigma$ is trivial cohomologycally.
\end{remark}

The cocycle $\sigma$ gives rise to a 2-cocycle on $\g$ with values
in $\RR$.

\begin{definition}
Let us consider the application $\Sigma:\g\times\g\rightarrow\RR$
given by ${\displaystyle \Sigma(\xi,\eta) =
<d\hat{\sigma}_{\eta}(e),\xi>}$, where $\hat{\sigma}_{\eta}:G
\rightarrow \RR$ is defined as $\hat{\sigma}_{\eta}(g) =
<\sigma(g),\eta>$.

\end{definition}

\begin{remark}

That $\Sigma$ is a skew-symmetric bilinear form on $\g$ satisfying
Jacobi's identity can be verified. If the trivial action of $\g$
on $\RR$ is considered, this last condition says that $\Sigma$ is
a 2-cocycle on $\g$ with values in $\RR$ associated to the trivial
action.

In addition, we can see that
$$\Sigma(\xi,\eta)=\{{\J}_0(\xi),{\J}_0(\eta)\}-{\J}_0([\xi,\eta]).$$

A $2$-cocycle is called exact if there exists $\mu \in {\g}^*$
such that $\Sigma(\xi,\zeta) = \mu ([\xi,\zeta])$. Then, is clear
that ${\J}_0$ is $Ad^*$-equivariant if and only if its associated
2-cocycle $\Sigma$ is trivial.

Thus, a non $Ad^*$-equivariant classical momentum map ${\J}_0$
canonically defines a central extension of ${\g}$ associated to
the $2$-cocycle $\Sigma$, which we will denote $\widetilde{\g}$.
This central extension will be considered in section \ref{section
6}.

\end{remark}

\section{Deformation Quantization and Fedossov's
construction}\label{section deformation quantization}

\hspace{0.4cm}In classical mechanics, phase spaces can be
represented by symplectic manifold $M$ and classical observables
are elements in the Poisson algebra $C^{\infty}(M)$.
 In quantum mechanics, observables form a noncommutative associative algebra.

Deformation quantization provides a mathematical framework for the
problem of quantizing classical mechanical systems. Its goal is to
construct and analyze noncommutative algebras corresponding to
Poisson algebras by means of formal deformations depending on the
Plank constant $\hbar$ as parameter \cite{Henry}.

To this end, start products that deform the point products of the
differential functions in the direction of a given bracket can be
constructed.

Now, let us consider a symplectic manifold $(M,\omega)$ and
$\{\cdot,\cdot\}$ the Lie algebra structure on $C^{\infty}(M)$
canonically associated to $\omega$.

\begin{definition}
 $C^{\infty}(M)[[\hbar]]$ is the vector space of formal power series with coefficients in $C^{\infty}(M)$. That is, $$C^{\infty}(M)[[\hbar]]=\left\{\d\sum_{r=0}^{\infty}f_r\hbar^r\ :\ f_r\in C^{\infty}(M)\ \forall\ r\geq 0\right\}.$$
 In general, the space $V[[\hbar]]$ can be consider for any vector space $V$,  $$V[[\hbar]]=\left\{\d\sum_{r=0}^{\infty}v_r\hbar^r\ :\ v_r\in V\ \forall\ r\geq 0\right\}.$$
\end{definition}

\begin{definition}{\label{producto estrella}}

A deformation quantization of $C^{\infty}(M)$ or a start product
$\star$ , is an associative algebra structure on
$C^{\infty}(M)[[\hbar]]$ of the form
$$f\star g=\sum^{\infty}_{r=0}C_{r}(f,g)\, \hbar^{r}\ \ \forall \ f\ \mbox{and}\ g\in
C^{\infty}(M)[[\hbar]],$$ where

% xu pag 1 y 2

\begin{enumerate}
\item $C_0(f,g)=f.g$ (pointwise multiplication);

\item $C_1(f,g) = - \frac {i}{2} \{f,g\}$

\item $C_r(f,g)=(-1)^r C_r(g,f)$;

\item $C_r(1,f)=C_r(f,1)$, for $r\geq 1$;

\item each $C_r(\cdot,\cdot)$ is a bidifferential operator.
\end{enumerate}

Then,

$$f\star g=f\, g - \frac {i\hbar}{2} \{f,g\} + \sum^{\infty}_{r=2}C_{r}(f,g)\, \hbar^{r}\
\ \ \forall \ f \ \mbox{and} \ g \in C^{\infty}(M).$$

\end{definition}

\begin{remark}
A start product $\star$ on $C^{\infty}(M)[[\hbar]])$ gives it a
Lie algebra structure by means of the bracket ${\displaystyle
[f,g]_\star = f \star g - g \star f}$. The adjoint representation
$(ad_\star)$ of $C^{\infty}(M)[[\hbar]]$ on itself is given by
${\displaystyle (ad_\star)_{\gamma}(\cdot)=[\gamma,\cdot]_\star}$.

\end{remark}
\begin{definition}
A derivation of a $\star$-algebra $(C^{\infty}(M)[[\hbar]],\star)$
is a formal power series of $\hbar$ with coefficients being linear
operators on $C^{\infty}(M)$, written as $\delta=D_0+\hbar
D_1+...+\hbar^i D_i+...$, such that $$\delta(f\star g)=\delta
f\star g+f\star\delta g\ \ \forall\ f\ \mbox{and}\ g\in
C^{\infty}(M)[[\hbar]].$$ A derivation is said to be inner if
$\delta(\cdot)=\frac{i}{\hbar}(ad_\star)_H(\cdot)=\frac{i}{\hbar}[H,\
\cdot\ ]_\star$ for some $H\in C^{\infty}(M)[[\hbar]]$.
\end{definition}
\ \\

\begin{example}
A star product always exists on a symplectic vector space $V$,
which is known as the Moyal-Weyl formula
$$f\circ g=\d\sum_{k=0}^{\infty}(-\frac{i\hbar}{2})^k\frac{1}{k!}
\pi^{i_1 j_1}...\pi^{i_k j_k}\frac{\partial^k f}{\partial
y^{i_1}...\partial y^{i_k}} \frac{\partial^k g}{\partial
y^{j_1}...\partial y^{j_k}}\ \ \forall \ f\ \mbox{and} \ g \in
C^{\infty}(V)[[\hbar]],$$ where $y^1,...,y^{2n}$ are linear
coordinates on V and $\pi^{ij}=\{y^i,y^j\}$.

The existence proof of star products on a general symplectic
manifold was first obtained by Wilde and Lecomte \cite{De
Wilde-Lecomte} using a homological argument.
\end{example}
\ \\

Now, we recall some basic ingredients of Fedosov construction of
$\star$-products on a symplectic manifold; consult \cite{Xu} and
\cite{alemanes} for details.

Let us consider
$$W\otimes\Lambda=\left(\d\prod_{s=0}^{\infty}\Gamma^{\infty}(\vee^sT^*M\otimes
\wedge T^*M)\right)[[\hbar]],$$ where $\Gamma^{\infty}(P)$ denotes
the space of sections of a vector bundle $P$ on $M$.
$W\otimes\Lambda$ becomes in a natural way an associative
super-commutative algebra and the product is denoted by
$\mu(a\otimes b)=ab$ for $a$ and $b\in W\otimes \Lambda$. By
$W\otimes\Lambda^k$ we denote the elements of anti-symmetric
degree $k$ and $W=W\otimes\Lambda^0$.

Let be $i_s(Y)$ and $i_a(Y)$ the symmetric and the anti-symmetric
insertion of a vector field $Y\in\Gamma^{\infty}(TM)$,
respectively .

Then, the pointwise product the Poisson tensor $\Lambda$
corresponding to $\omega$ gives rise to another associative
product $\circ$ on $W\otimes\Lambda$ by $$a\circ b=\mu\circ exp\
(\d\frac{1}{2}\hbar\Lambda^{ij}i_s(\partial_i)\otimes
i_s(\partial_j))(a\otimes b),$$ which is a deformation of $\mu$.

Sometimes we write $W_k\otimes\Lambda$ to denote the elements of
total degree $\geq k$.

In local coordinates they define the differential
$\delta=(1\otimes dx^i)_{i_s}(\partial_i)$ which satisfies
$\delta^2=0$ and is a super-derivation of $\circ$. There is a
homotopy operator $\delta^{-1}$ satisfying
$\delta\delta^{-1}+\delta^{-1}\delta+\sigma=id$ where
$\sigma:W\otimes\Lambda\longrightarrow C^{\infty}(M)[[\hbar]]$
denotes the projection onto the part of symmetric and
anti-symmetric degree $0$ and
$\delta^{-1}a=\frac{1}{k+l}(dxî\otimes 1)i_a(\partial_i)a$ for
$deg_sa=ka$, $deg_aa=la$ with $k+l\neq 0$ and $\delta^{-1}a=0$,
where $deg_a$ and $deg_s$ are de obvious degree maps.

From a torsion free symplectic connection $\nabla$ on $M$ we
obtain a derivation of $\circ$ $$\nabla=(1\otimes
dx^i)\nabla_{\partial_i}$$ that satisfies the following identities
$$[\delta,\nabla]=0,\ \ \ \nabla^2=-(\frac{1}{\hbar})ad(R)$$ where
$R=\frac{1}{4}\omega_{it}R^t_{jkl}dx^i\vee dx^j\otimes dx^k\wedge
dx^l\in W\otimes\wedge^2$ involves the curvature of the
connection.

Now remember the following facts which are just restatements of
Fedosov's original theorems.

For all $\Omega\in \hbar Z^2_{dR}(M)[[\hbar]]$ and all $s\in W_3$
with $\sigma(s)=0$ there exists a unique element $r\in
W_2\otimes\Lambda^1$ such that
\begin{equation}\label{delta r}
\delta r=\nabla r-\frac{1}{\hbar}r\circ r+R+1\otimes\Omega\ \
 \mbox{and}\ \ \delta^{-1}r=s.
\end{equation}
Moreover $r$ satisfies the formula $r=\delta s+\delta^{-1}(\nabla
r-\frac{1}{\hbar}r\circ r+R+1\otimes\Omega)$ from which $r$ can be
determined recursively. In this case the Fedosov derivation
$$D=-\delta+\nabla-\frac{1}{\hbar}ad_r$$ is a super-derivation of
anti-symmetric degree $1$ and $D^2=0$. The connection $D$ defined
above is a Fedosov connection corresponding to
$(\nabla,\Omega,s)$.

For any $f\in C^{\infty}(M)[[\hbar]]$ there exists a unique
element $\tau(f)\in ker(D)\cap W$ such that $\sigma(r(f))=f$ and
$\tau:C^{\infty}(M)[[\hbar]]\longrightarrow ker(D)\cap W$ is
$\mathbb{C}[[\hbar]]$-linear and it is referred to as the
Fedosov-Taylor series corresponding to $D$. In addition $\tau(f)$
can be obtained recursively for $f\in C^{\infty}(M)$ from
$$\tau(f)=f+\delta^{-1}\left(\nabla\tau(f)-\frac{1}{\hbar}ad_r\tau(f)\right).$$

Using $D^{-1}$ one can also write $\tau(f)=f-D^{-1}(1\otimes df)$.
$D$ constructed as above is a $\circ$-super-derivation, $ker
(D)\cap W$ is a $\circ$-sub-algebra and a new associative product
$\star$ for $C^{\infty}(M)[[\hbar]]$, which turns out to be a star
product, is defined by pull-back of $\circ$ via $\tau$.

Observe that in (\ref{delta r}) we allowed for an arbitrary
element $s\in W$ with $\sigma(s)=0$ that contains no term of total
degree lower than $3$ as normalization condition for $r$, i.e.
$\sigma^{-1}r=s$ instead of the usual equation $\delta^{-1}r=0$.

In the following we shall refer to the associative product $\star$
defined above as the Fedosov star product corresponding to
$(\nabla,\Omega,s)$.

Besides, as an important consequence of some propositions
(\cite{alemanes}) they get the following consequence.

For every Fedosov star product $\star$ obtained from
$(\nabla,\Omega,s)$ with $s\in W_3$ there is a connection
$\nabla'$, a formal series $\Omega'$ of closed two-forms and an
element $s'\in W_4$ without terms of symmetric degree $1$ such
that the star product obtained from $(\nabla',\Omega',s')$
coincides with $*$, and hence we may without loss of generality
restrict to such normalization conditions when varying the
connection and the formal series of closed two-forms arbitrarily.
\ \\

Now, let us consider, as in \cite{alemanes}, a generalized
Fedosov's  $\star$ product on a symplectic manifold $(M,\omega)$
stemming from a Weyl product $\circ$ and a triad
$(\nabla,\Omega,s)$ constituted by a flat torsion free symplectic
connection $\nabla$ on $M$, a 2-form formal series $\Omega$ on $M$
and a symmetrical contravariant tensors formal series $s\in W_4$
on $M$ contains no part of symmetric degree $1$.

As in the previous section, we consider a Lie group $G$ that
symplectically acts in $(M,\omega)$. The action $\rho$ of ${\g}$
on $C^{\infty}(M)$ defined above can be naturally extended to an
action $\rho_c$ of ${\g}$ on $C^{\infty}(M)[[\hbar]]$ in the
following way

$$\rho_{c}(\xi)\left(\sum_{r\geq 0}f_{r}\hbar^{r}\right)=-\sum_{r\geq 0}(L_{X_{\xi}}f_{r})\,
\hbar^{r}.$$ Thus, $C^{\infty}(M)[[\hbar]]$ becomes in a
${\g}$-module.

We shall consider the cohomolgy of ${\g}$ with coefficients in
$C^{\infty}(M)[[\hbar]]$. Its coboundary operator

$$\delta_{\rho_c}:C^{k}({\g},C^{\infty}(M)[[\hbar]]) \rightarrow C^{k+1}({\g},C^{\infty}(M)
[[\hbar]]) \ \ \ \forall \ k \in \ \NN$$ is defined as in the
previous section.

\begin{definition}

A star product $\star$ is called ${\g}$-invariant in case
$\rho_c(\xi)$ is a derivation of $\star$ for all $\xi\in{\g}$.
That is,
$$\rho_{c}(\xi)(f\star g)=(\rho_{c} (\xi)(f)\star g)+(f\star\rho_{c} (\xi)(g)) \ \ \forall \ f, g \in C^{\infty}(M)[[\hbar]] \
\mbox{and} \ \forall \ \xi \in \g.$$

\end{definition}

On \cite{alemanes}, the following characterization of these
products can be found.

\begin{proposition}

The Fedosov star product $\star$ constructed from
$(\nabla,\Omega,s)$, where $s\in W_4$ contains no part of
symmetric degree $1$, is $\g$-invariant if and only if $X_{\xi}$
is affine with respect to $\nabla$ for all $\xi\in\g$, that is $$
[\nabla, L_{X_ {\xi}} ] =0\ \ \forall\ \xi\in \g$$ and $\Omega$
and $s$ are invariant with respect to $X_{\xi}$ for all
$\xi\in\g$, that is $$d\, i_{X_{\xi}} \Omega = L_{X_ {\xi}} \Omega
= 0=L_{X_{\xi}}s \ \ \forall \ \xi \in \g.$$
\end{proposition}

In the next section these star products will be considered.

\ \\

\section{Quantum Momentum Map}

\hspace{0.4cm} Now, we shall recall the definition of a quantum
momentum map and some of its properties given in \cite{alemanes}.

\begin{definition}

Given $\star$  a $\g$-invariant star product, ${\J} \in
C^{1}({\g},C^{\infty}(M))[[\hbar]]$ is called quantum Hamiltonian
for the action $\rho_{c}$ if

\begin{equation}\label{def. de hamiltoniano cuantico}
\rho_{c}(\xi)(\cdot)=\frac{1}{\hbar}(ad_{\star})_{{\J}(\xi)}(\cdot)
= \frac{1}{\hbar} [{\bf J}(\xi),\ \cdot\ ]_\star\ \ \ \forall \
\xi\in{\g}.
\end{equation}

That is, $\rho_c(\xi)$ is an inner derivation for all
$\xi\in{\g}$.

A quantum Hamiltonian $\J$ is a quantum momentum map if
${\J}:{\g}\longrightarrow
(C^{\infty}(M)[[\hbar]],\frac{1}{\hbar}[\cdot,\cdot]_\star)$ is a
Lie algebra homomorphism. That is,
\begin{equation}\label{def. de momento cuantico}
\frac{1}{\hbar}\left({\J}(\xi)\star
{\J}(\eta)-{\J}(\eta)\star{\J}(\xi)\right)= {\J}([\xi,\eta]) \ \
\forall \ \xi \ \mbox{and} \ \eta \in {\bf g}.
\end{equation}

\end{definition}
\ \\

\begin{remark}\label{remark 4.2}

Notice that ${\J}$ can be written as

$${\J} = {\J}_0 + {\bf J_+} \ \ \ \mbox{where} \ \ \ {\J}_0 \in C^{1}({\g},C^{\infty}(M))
 \ \ \mbox{and} \ \ \
{\bf J_+} \in \hbar \, C^{1}({\g},C^{\infty}(M))[[\hbar]].$$

It is very simple to see that the zeroth order in $\hbar$ of
(\ref{def. de hamiltoniano cuantico}) is equivalent to ${\J}_0$
being a classical momentum map and that the zeroth order in
$\hbar$ of (\ref{def. de momento cuantico}) just means
$Ad^*$-equivariance of this classical momentum map with respect to
the coadjoint action of $\g$.

Thus, it is clear that a quantum momentum map always gives rise to
an $Ad^*$-equivariant classical momentum map when the deformation
parameter $\hbar$ tends to $0$.
\end{remark}
\ \\

In the first place, we consider the necessary and sufficient
conditions for the existence of a quantum Hamiltonian that have
been proven in \cite{alemanes}.

\begin{proposition}

A ${\g}$-invariant Fedosov star product for $(M,\omega)$ obtained
from $(\nabla,\Omega,s)$ admits a quantum Hamiltonian if and only
if there is an element ${\J}\in
C^{1}({\g},C^{\infty}(M))[[\hbar]]$ such that
$$d{\J}(\xi)=i_{X_{\xi}}(\omega+\Omega)\ \ \forall \ \xi\in{\g}.$$
\end{proposition}
\ \\

In order to establish the necessary and sufficient conditions for
the existence of a quantum momentum map in \cite{alemanes} the
following result has been proven.

\begin{proposition}\label{proposicion 4.4}

Let ${\J}$ be a quantum Hamiltonian for the Fedosov star product
$\star$. Then,
$${\bf {\lambda}}\in C^{2}({\g},C^{\infty}(M))[[\hbar]] \
\mbox{defined by} \ {\bf
{\lambda}}(\xi,\eta)=\frac{1}{\hbar}\left({\J}(\xi)\star{\J}(\eta)-{\J}(\eta)\star{\J}(\xi)\right)
-{\J}([\xi,\eta])$$  is an element of $Z^{2}({\g},\RR)[[\hbar]]$
which is explicitly given by
$${\bf {\lambda}}(\xi,\eta)=(\omega+\Omega)(X_{\xi},X_{\eta})-{\J}([\xi,\eta]).$$
\end{proposition}

\begin{corollary}\label{corolario 4.5}
Let $\star$ be a $\g$-invariant Fedosov star product and let us
assume that there is an $Ad^*$-equivariant classical momentum map
${\J}_0$. Then, there exists ${\J}\in
({\g},C^{\infty}(M))[[\hbar]]$ quantum momentum map that recovers
${\J}_0$ as its classical limit if and only if there exists
${\J}_{+} \in \hbar.C^{1}({\g},C^{\infty}(M))[[\hbar]]$ such that
$$i_{X_{\xi}}\Omega=d\, {\J}_{+}(\xi)\ \ \forall\ \xi \in g$$ and
 $$\Omega(X_{\xi},X_{\eta})=(\delta_{\rho_c}{\J}_{+})(\xi,\eta)\ \ \forall\ \xi\ \mbox{and}\ \eta
 \in \g.$$

\end{corollary}

\begin{remark}

In this case ${\J}$ can be written as ${\J}={\J}_0+{\J}_+$.

Let us observe that
$\Omega(X_{\xi},X_{\eta})=(\delta_{\rho_c}{\J}_{+})(\xi,\eta)$ can
be written as
$$\Omega(X_{\xi},X_{\eta})= -L_{X_{\xi}}{\J}_{+}(\eta) + L_{X_{\eta}}{\J}_{+}(\xi) -
 {\J}_{+}([\xi,\eta]).$$
\end{remark}
\bigskip

\section{Anomalous Quantum Momentum Maps}

\hspace{0.4cm} As we said in remark \ref{remark 4.2}, the
condition (\ref{def. de hamiltoniano cuantico}) of the definition
of a quantum momentum map describes the condition for a quantum
counterpart of a classical momentum map. Meanwhile, the condition
(\ref{def. de momento cuantico}) corresponds to the property of
$Ad^*$-equivariance of a classical momentum map.

Let us notice that a non $Ad^*$-equivariant classical momentum map
cannot be considered as the classical limit of a quantum momentum
map but can be recovered from a quantum Hamiltonian. Thus, a
quantum Hamiltonian
${\J}:{\g}\longrightarrow(C^{\infty}(M)[[\hbar]],\frac{1}{\hbar}[\cdot,\cdot
]_\star)$ such that is not a Lie algebra homomorphism can be
considered as the quantum counterpart of a non $Ad^*$-equivariant
classical momentum map.

\begin{definition}
An anomalous quantum momentum map is a quantum Hamiltonian
${\J}:{\g}\longrightarrow(C^{\infty}(M)[[\hbar]],\frac{1}{\hbar}[\cdot,\cdot
]_\star)$ such that is not a Lie algebra homomorphism.
\end{definition}

Then, if ${\J}$ is an anomalous quantum momentum map, the
condition $$\d\frac{1}{\hbar}\left({\J}(\xi)\star
{\J}(\eta)-{\J}(\eta)\star{\J}(\xi)\right)={\J}([\xi,\eta])$$ is
not fulfilled. That is the $2$-cocycle
$$\lambda(\xi,\eta)=\d\frac{1}{\hbar}\left({\J}(\xi)\star
{\J}(\eta)-{\J}(\eta)\star{\J}(\xi)\right)-{\J}([\xi,\eta])$$
results not null.

\begin{remark}
It is obvious that an anomalous quantum momentum map is a quantum
momentum map if and only if the 2-cocycle ${\bf {\lambda}}\in
C^{2}({\g},\RR)[[\hbar]]$ is trivial. In this case, the anomalous
quantum momentum map recovers an $Ad^*$-equivariant classical
momentum map.
\end{remark}

Let us consider a non $Ad^*$-equivariant classical momentum map
${\J}_0$. As a consequence of Corollary \ref{corolario 4.5} it is
easy to characterize the existence of an anomalous quantum
momentum map that recovers ${\J}_0$.

\begin{proposition}\label{existencia de momento anomalo}

There exists ${\J} = {\J}_0 + {\J}_+$ an anomalous quantum
momentum map if and only if there exists $\ {\J}_+ \in \hbar \,
C^{1}({\g},C^{\infty}(M))[[\hbar]]$ such that $i_{X_{\xi}} \Omega
= d\, {\J}_+(\xi)$.

\end{proposition}

In the following sections, we shall define two quantum momentum
maps on different central extensions of ${\g}$. We shall see their
restrictions on ${\g}$ are anomalous quantum momentum maps that
recover the non  $Ad^*$-equivariant classical momentum map
${\J}_0$.

\section{Quantum Momentum Map Associated to the Canonically Extended Classical Momentum Map}\label{section 6}

\subsection{Canonical Extended Classical Momentum Map $\widetilde{\J}_0$}

\hspace{0.4cm}As we remark in the Subsection \ref{subsection 2.2},
in the case of a symmetry with a non $Ad^*$-equivariant momentum
map ${\J}_0$, there exists a canonical central extension of the
Lie algebra ${\g}$ given by the cocycle that measures the non
$Ad^*$-equivariance of the momentum map.

\begin{definition}

Let $\gtil$ the central extension of the Lie algebra ${\bf g}$
associated to the $2$-cocycle $\Sigma$. Then, $\gtil = \g \oplus
\RR$ and $\gtil^* = \g^* \oplus \RR$.

The Lie commutator on $\gtil$ is given by
$$[(\xi,a),(\eta,b)]=([\xi,\eta],\Sigma(\xi,\eta)) \ \ \ \forall \ (\xi,a) \ \ \mbox{and} \ \ \ (\eta,b)
\in \gtil$$ and the evaluation of an element of $\gtil^*$ on
$\gtil$ is defined as
 $$\langle (\alpha,x),(\xi,a) \rangle = \alpha(\xi) +
x\, a \ \ \ \forall \ (\alpha,x)
 \in \gtil^* \ \ \mbox{and} \ \ (\xi,a) \in \gtil$$
\end{definition}

\begin{remark}

We shall assume that the extension $\gtil$ of the Lie algebra $\g$
corresponds to an extension $\widetilde{G}$ of the Lie group $G$.

Let us consider the trivially extended action of $\widetilde{G}$
on $M$, $\fit:\widetilde{G}\times M\rightarrow M$ defined as
$$\displaystyle \fit_{(g,A)}(m)=\phi_{g}(m)\ \ \forall\ (g,A) \in
\widetilde{G}\ \mbox{and}\ m \in M.$$

Then the infinitesimal generator $\Xtil_{(\xi,a)}$ associated to
$(\xi,a) \in \gtil$ coincides with the infinitesimal generator
$X_{\xi} \ \forall \ \xi \in {\bf g}.$

\end{remark}

In a canonical way, we can define a representation
$\widetilde{\rho}$ of $\gtil$ in $C^{\infty}(M)$ given by
\begin{eqnarray*}\widetilde{\rho} : \widetilde{\g} \times C^{\infty}(M)
& \rightarrow & C^{\infty}(M) \\
\widetilde{\rho}(\xi,a)(f) & = & - L_{\widetilde{X}_{(\xi,a)}}f
\end{eqnarray*}
and its associated coboundary operator
$\delta_{\widetilde{\rho}}$.

From the non $Ad^*$-equivariant momentum map we can define a
momentum map that results $Ad^*$-equivariant with respect to an
extended coadjoint action of $\gtil^*$ defined as
$$\widetilde{Ad}^*_{(g,A)}(\alpha,x)=(Ad^*_{g}\alpha+\sigma(g),x)\ \ \forall\ (\al,x)\in\widetilde{\g}^*.$$
\ \\

Let us consider the applications

\begin{itemize}
\item $\Jotil:M\rightarrow \gtil^* \ \mbox{given by} \
\Jotil(m)=(\Jo(m),1)$ and \item
$\widetilde{{\J}}_0:\gtil\rightarrow C^{\infty}(M) \ \mbox{defined
as} \ \widetilde{\J}_0(\xi,a)(m)={\J}_0(\xi)(m)+a$
\end{itemize}

Its easy to see that $d\, \widetilde{\J}_0(\xi,a) =
i_{\Xtil_{(\xi,a)}} \omega$. Thus, we can define an extended
classical momentum map as follows.

\begin{definition}

The applications  $\Jotil$ and $\widetilde{{\J}}_0$ are called
extended classical momentum maps associated to the action
$\widetilde{\phi}$ of $\widetilde{G}$ on $M$.

\end{definition}

It is clear that $\widetilde{{\J}}_0 \in
C^1(\widetilde{\g},C^{\infty}(M))$.

\ \\

Next we shall see that this extended momentum map
$\widetilde{{\J}}_0$ is $\widetilde{Ad}^*$-equivariant.

\begin{proposition}
$\widetilde{\J}_0:\gtil\rightarrow C^{\infty}(M)$ is a
 Lie algebra homomorphism.

\end{proposition}

{\bf Proof.}

$$\begin{array}{rcl}
\{\widetilde{\J}_0(\xi,a),\widetilde{\J}_0(\eta,b)\}&=&\{{\J}_0(\xi)+a,{\J}_0(\eta)+b\}\vspace{0.2cm}\\
                &=&\{{\J}_0(\xi),{\J}_0(\eta)\}+\{{\J}_0(\xi),b\}+\{a,{\J}_0(\eta)\}+\{a,b\}\vspace{0.2cm}\\
                &=&{\J}_0([\xi,\eta])+\Sigma(\xi,\eta)\vspace{0.2cm}\\
                &=&\widetilde{\J}_0([\xi,\eta],\Sigma(\xi,\eta))\vspace{0.2cm}\\
                &=&\widetilde{\J}_0([(\xi,a),(\eta,b)]),
\end{array}$$
for all $(\xi,a)$ and $(\eta,b)\in \widetilde{\g}$.\ $\triangle$

\ \\

\subsection{Quantum Momentum Map associated to $\widetilde{\J}_0$}

\hspace{0.4cm}According to \cite{alemanes}, we shall study the
existence of a quantum momentum map associated to the extended
classical momentum map $\widetilde{\J}_0$.

In the first place, let us notice that the representation
$\widetilde{\rho}$ of $\gtil$ on $C^{\infty}(M)$ can be
canonically extended to the space $C^{\infty}(M)[[\hbar]]$ as
follows. Let $\widetilde{\rho}_{c}:\gtil\times
C^{\infty}(M)[[\hbar]]\rightarrow C^{\infty}(M)[[\hbar]]$ be given
by
$$\rhotil_{c}(\xi,a)\left(\sum_{r\geq 0}f_{r}\hbar^{r}\right)=
\sum_{r\geq 0}\rhotil(\xi,a)(f_{r})\hbar^{r}=\sum_{r \geq
0}\rho(\xi)(f_{r})\hbar^{r}.$$

We denote as $\delta_{\rhotil_{c}}$ its coboundary operator
associated.

In order to assign a quantum momentum map to $\widetilde{\J}_0$,
we shall consider a $\gtil-\mbox{invariant}$ Fedosov star product.
It is easy to verify that this is equivalent to considering a
$\g-\mbox{invariant}$ Fedosov star product.

\begin{proposition}
A Fedosov star product $\star$ is $\gtil-\mbox{invariant}$ if and
only if it is a $\g-\mbox{invariant}$ Fedosov star product.
\end{proposition}

{\bf Proof.}

A Fedosov star product $\star$ is $\gtil$-invariant if and only if

$$\rhotil_c(\xi,a)(f\star g)=(\rhotil_c(\xi,a)(f))\star g + f\star(\rhotil_c(\xi,a)(g)).$$

Since $\rhotil_c(\xi,a)(f) = -L_{\Xtil_{(\xi,a)}}f = -L_{X_{\xi}}f
= \rho_c(\xi)(f)$, then the above condition is equivalent to
$\rho(\xi)(f\star g)=\left( \rho(\xi)(f)\star g \right)+\left(
f\star\rho(\xi)(g) \right)$. That is, $\star$ is $\g$-invariant.
$\bigtriangleup$.
\ \\

Now, let us consider $\star$ a ${\bf g}$-invariant Fedosov start
product defined from $(\nabla,\Omega,s)$ as has been described in
section \ref{section deformation quantization}.

Following \cite{alemanes} we consider a quantum momentum map
corresponding to the representation $\rhotil_c$ and establish the
necessary and sufficient conditions for its existence. \ \\

\begin{remark}
The application ${\bf \Jtil} = \widetilde{{\J}}_0 + {\bf\Jtil_{+}}
:\gtil\rightarrow C^{\infty}(M)[[\hbar]]$ with ${\bf \Jtil}_0 \in
C^1(\widetilde{\g},C^{\infty}(M))$ and $\widetilde{\J}_+ \in
\hbar\, C^1(\widetilde{\g}, C^{\infty}(M))[[\hbar]] $ is a quantum
momentum map for $\rhotil_c$ if and only if

\begin{enumerate}
\item $\rhotil_{c}(\xi,a)(\cdot)=\d\frac{1}{\hbar}\
ad_\star({\bf\Jtil}(\xi,a))(\cdot) = \d\frac{1}{\hbar}\left[{\bf
\Jtil}(\xi,a),\ \cdot\ \right]_\star \ \ \ \f\ (\xi,a)\in\gtil$

\item
${\bf\Jtil}\left([(\xi,a),(\eta,b)]\right)=\d\frac{1}{\hbar}\left[{\bf
\Jtil}(\xi,a),{\bf \Jtil}(\eta,b)\right]_\star\ \ \forall\
(\xi,a)$ and $(\eta,b)\in\widetilde{\g}$.
\end{enumerate}
\end{remark}

From the corollary (\ref{corolario 4.5}) we can establish the
following result.

\begin{corollary}\label{existencia de momento quantico extendido}
${\bf \Jtil}={\bf \Jtil}_0+{\bf \Jtil_{+}}:\gtil\rightarrow
C^{\infty}(M)[[\hbar]]$ is a quantum momentum map for
$\rhotil_{c}$ if and only if there exists ${\bf \Jtil_{+}}\in\hbar
\ C^{1}(\gtil,C^{\infty}(M))[[\hbar]]$ such that
$$i_{\Xtil_{(\xi,a)}}\Omega=d{\bf \Jtil_{+}}(\xi,a)$$ and
$$\Omega\left(\Xtil_{(\xi,a)},
\Xtil_{(\eta,b)}\right)=\left(\delta_{\rhotil_{c}}{\bf
\Jtil_{+}}\right)_{((\xi,a),(\eta,b))}.$$
\end{corollary}
\ \\

Thus, we can recover ${\J}_0$ as the classical limit of the
restriction of $\widetilde{\J}$ to ${\g}$.

\begin{proposition}
$\breve{\J}=\widetilde{\J}|_{{\g}\oplus\{0\}}$ is an anomalous
quantum momentum map that recovers the non $Ad^*$-equivariant
classical momentum map ${\J}_0$.
\end{proposition}

{\bf Proof.}

It is clear that $\breve{\J}$ is a quantum Hamiltonian,

\begin{equation}\label{J brebe es quantum hamiltonian}
\rho_c(\xi)(\cdot)=\widetilde{\rho}_c(\xi,0)(\cdot)=\frac{1}{\hbar}[\widetilde{\J}(\xi,0),\
\cdot\ ]_\star=\frac{1}{\hbar}[\breve{\J}(\xi),\ \cdot\ ]_\star.
\end{equation}

Also we can see that $\breve{\J}$ is not a Lie algebra
homomorphisms,

$$[\breve{\J}(\xi),\breve{\J}(\eta)]_\star=[\widetilde{\J}(\xi,0),\widetilde{\J}(\eta,0)]_\star=\widetilde{\J}([(\xi,0),(\eta,0)])=\widetilde{\J}([(\xi,\eta)],\Sigma(\xi,\eta)),$$

and $$\breve{\J}([\xi,\eta])=\widetilde{\J}([\xi,\eta],0).$$

Then $\breve{\J}$ is a quantum anomalous momentum map.

On the other hand,
$$\breve{\J}(\xi)=\widetilde{\J}(\xi,0)=\widetilde{\J}_0(\xi,0)+\widetilde{\J}_+(\xi,0)={\J}_0(\xi)+0+\widetilde{\J}_+(\xi,0).$$

Then it is clear that $\breve{\J}$ recovers ${\J}_0$ when $\hbar$
tends to zero.\ $\triangle$

\begin{remark}
Let us notice that ${\J}_0$ can be recovered as classical limit of
the restriction of $\widetilde{\J}$ to ${\g}$ if and only if there
exists $\widetilde{\J}_+\in\hbar \
C^{1}(\gtil,C^{\infty}(M))[[\hbar]]$ such that verifies the
conditions of Corollary (\ref{existencia de momento quantico
extendido}).
\end{remark}
\section{Canonical Extended Anomalous Quantum Momentum Maps}

\ \\

Given a non $Ad^*$-equivariant classical momentum map ${\J_0}$, as
we have established in Proposition \ref{existencia de momento
anomalo}, there exists an anomalous quantum momentum map ${\bf J}
= {\bf J}_{0} + {\bf J}_{+}$ if and only if there exists $\ {\bf
J_+} \in \hbar \, C^{1}({\g},C^{\infty}(M))[[\hbar]]$ such that
$i_{X_{\xi}} \Omega = d\, {\bf J_+}(\xi)$.\ \\

Let us assume that such ${\J}$ there exists. In order to define a
quantum momentum map on a central extension of ${\g}$ whose
restriction to ${\g}$ coincide whit ${\J}$, we analyze the
2-cocycle ${\bf \lambda}: {\g} \times {\g} \rightarrow
\RR[[\hbar]]$ given by
$${\bf \lambda}(\xi,\eta)=\d\frac{1}{\hbar}\left[ {\J}(\xi),
{\J}(\eta)\right]_\star-{\J}([\xi,\eta]).$$

Notice that ${\bf \lambda}$ can be written as ${\bf
\lambda}(\xi,\eta)=\d\sum_{r\geq 0}\lambda_{r} (\xi,\eta)\,
\hbar^{r}$, where $\lambda_{r}\in C^{2}({\g},\RR)$ for all $r \in
\RR$. If $\hbar$ tends to zero, we obtain that
$\displaystyle{\lambda_{0}(\xi,\eta)=\{{\J}_0(\xi),{\J}_0(\eta)\}-{\J}_0([\xi,\eta])=\Sigma(\xi,\eta).}$

It is clear that $\RR[[\hbar]]$ becomes a ${\g}$-module by the
trivial action. Thus, we can consider the central extension
${\gsom}$ of ${\g}$ by $\RR[[\hbar]]$ associated to the 2-cocycle
${\bf \lambda}$.

Then, ${\displaystyle {\gsom} = {\g}\oplus\RR[[\hbar]]}$ and the
bracket in $\gsom$ is given by
$$\left[ \left(\xi,\sum_{r\geq 0}x_{r}\hbar^{r}\right),\left(\eta,\sum_{r\geq
0}y_{r}\hbar^{r}\right)\right] = \left([\xi,\eta],{\bf
\lambda}(\xi,\eta)\right)$$ where $\xi,\eta \in {\g} \ \mbox{and}
\ \d\sum_{r\geq 0} x_{r}\hbar^{r} \ \mbox{and} \ \d\sum_{r\geq
0}y_{r}\hbar^{r} \in \RR[[\hbar]].$

As in the case of the central extension $\gtil$, ${\gsom}$ acts
trivially in the extension component on $C^{\infty}(M)$ and this
action is canonically extended on $C^{\infty}(M)[[\hbar]]$. That
is, $\widehat{\rho}: \gsom \times C^{\infty}(M)\rightarrow
C^{\infty}(M)$ is defined as

$$\widehat{\rho}\left(\xi,\d\sum_{r\geq 0} x_r\, {\hbar}^r\right)(f)=-L_{X_{\xi}}f,$$
and $\widehat{\rho}_c: \gsom \times
C^{\infty}(M)[[\hbar]]\rightarrow C^{\infty}(M)[[\hbar]]$ is given
by $$\widehat{\rho}_c\left(\xi,\d\sum_{r\geq 0} x_r\,
{\hbar}^r\right)\left(\d\sum _{r \geq 0} f_r\, {\hbar}^r\right)=-
\d\sum _{r \geq 0} (L_{X_{\xi}}f_r){\hbar}^r.$$

Next, we will define a quantum momentum map that canonically
extends $\bf J$.

Let us consider the application
 ${\displaystyle {\bf \Jsom}:\gsom\rightarrow
C^{\infty}(M)[[\hbar]]}$ given by
$${\bf \Jsom}\left(\xi,\sum_{r\geq
0}x_{r}\hbar^{r}\right) = {\J}(\xi) + \sum_{r\geq
0}x_{r}\hbar^{r}$$

In the first place, we notice that it is easy to verify that

$$\widehat{\rho}_c \left(\xi, \d\sum_{r\geq 0} x_{r}\hbar^r\right)(\cdot) = \frac
{1}{\hbar}[\widehat{ {\J}}(\xi),\ \cdot\ ]_\star\ \ \ \forall\
\left(\xi, \d\sum_{r\geq 0} x_{r}\hbar^r\right) \in \gsom.$$

Then, $\widehat{\J}$ is a quantum Hamiltonian for the quantum
action $\widehat{\rho}_c$.
\ \\

In order to see that $\widehat{\J}$ is a Lie algebra homomorphism
we will prove the following lemma.

\begin{lemma}
If $f \in\RR[[\hbar]]$ then $[f,g]_\star=0$, for all $g\in
C^{\infty}(M)[[\hbar]]$.
\end{lemma}

{\bf Proof.}

Let $f\in\RR[[\hbar]]$ and $g\in C^{\infty}(M)[[\hbar]]$. Then
$f=\d\sum_{r\geq 0}a_r\hbar^r$, $g=\d\sum_{r\geq 0}g_r\hbar^r$
where $a_r\in\RR$ and $g_r\in C^{\infty}(M)$ for all $r\geq 0$.
Thus,

$$[f,g]_\star=\left[\d\sum_{r\geq 0}a_r\hbar^r,\d\sum_{r\geq
0}g_r\hbar^r\right]_\star=\d\sum_{r\geq
0}\left(\d\sum_{i+j=r}[a_i,g_i]_\star\right)\hbar^r=0,$$ where the
last equality is fulfilled because $[a,g]_\star=0$ for all
$a\in\RR$ and $g\in C^{\infty}(M)$. $\bigtriangleup$

\begin{proposition}
$\widehat{\J}:\widehat{\g}\longrightarrow(C^{\infty}(M)[[\hbar]],\frac{1}{\hbar}[\cdot,\cdot]_\star)$
is a Lie algebra homomorphism.
\end{proposition}

{\bf Proof.}

By definition of $\widehat{\J}$ and definition of the bracket in
$\widehat{\g}$,
$$\begin{array}{rcl}
\widehat{\J} \left(\left[(\xi,\d\sum_{r\geq 0}x_{r}\hbar^r),
(\eta,\d\sum_{r\geq
0}y_{r}\hbar^r)\right]\right)&=&\widehat{\J}([\xi,\eta],{\bf
\lambda}
(\xi,\eta))\\
&=&{\J}([\xi,\eta])+{\bf \lambda}(\xi,\eta).
\end{array}$$

 On the other hand,

 $$\begin{array}{rcl}
\d\frac{1}{\hbar}\left[\widehat{\J}\left(\xi,\d\sum_{r\geq 0}
x_{r}\hbar^r\right),\widehat{\J}\left(\eta,\d\sum_{r\geq 0}
y_{r}\hbar^r\right)\right]_\star&=&\d\frac{1}{\hbar}\left[{\J}(\xi)+\d\sum_{r\geq
0} x_{r}\hbar^r,{\J}(\eta)+\d\sum_{r\geq 0}
y_{r}\hbar^r\right]_\star\vspace{0.3cm}\\
&=&\d\frac{1}{\hbar}\left[{\J}(\xi),{\J}(\eta)\right]_\star+\d\frac{1}{\hbar}\left[{\J}(\xi),\d\sum_{r\geq 0} y_{r}\hbar^r\right]_\star\vspace{0.3cm}\\
&&+\d\frac{1}{\hbar}\left[\d\sum_{r\geq 0}
x_r\hbar^r,{\J}(\eta)\right]_\star+\d\frac{1}{\hbar}\left[\d\sum_{r\geq
0}
x_r\hbar^r,\d\sum_{r\geq 0} y_r\hbar^r\right]_\star\vspace{0.3cm}\\
&=&\d\frac{1}{\hbar}[{\J}(\xi),{\J}(\eta)]_\star
 \end{array}$$

This last equality is fulfilled because the three last brackets
are equal to zero. Then, by definition of ${\bf \lambda}$, it is
clear that $\widehat{\J}$ is a Lie algebra
homomorphism.$\bigtriangleup$
\ \\

Thus, the following is a good definition.

\begin{definition}

The application $\widehat{\J}:\gsom\longrightarrow
(C^{\infty}(M)[[\hbar]],\frac{1}{\hbar}[\cdot,\cdot]_\star)$ given
by
$${\bf \widehat{\J}}\left(\xi,\sum_{r\geq
0}x_{r}\hbar^{r}\right) = {\J}(\xi) + \sum_{r\geq
0}x_{r}\hbar^{r}$$ is a quantum momentum map corresponding to the
quantum action $\widehat{\rho}_c$.

$\widehat{\J}$ is called the canonical extended quantum momentum
map associated to ${\J}$.
\end{definition}
\ \\

We want to remark that $\widehat{\J}$ is a Lie algebra
homomorphism that gives rise a good application on
$\widetilde{\g}$ to $C^{\infty}(M)$ when $\hbar$ tends to zero.
The application can be written as

$$\begin{array}{rcl}
\widehat{\J}\left(\xi,\d\sum_{r\geq
0}x_r\hbar^r\right)&=&{\J}(\xi)+\d\sum_{r\geq 0}x_r\hbar^r\\
&=&{\J}_0(\xi)+{\J}_+(\xi)+\d\sum_{r\geq 0}x_r\hbar^r.
\end{array}$$

So, the classical limit of ${\J}_0(\xi)+{\J}_+(\xi)+\d\sum_{r\geq
0}x_r\hbar^r$ agree with $\widetilde{\J}(\xi,x_0)={\J}_0(\xi)+x_0$
for all $\xi\in{\g}$ and $x_0\in\RR$.

The process to take classical limit can be thought as to take
quotient by the ideal generated by $\hbar$ denoted by $[[\hbar]]$.
Then, we can consider the following diagram

$$\begin{array}{ccc}
\widehat{\g}              & \stk{\widehat{\J}}& C^{\infty}(M)[[\hbar]]\\
\dwn{q_{\hbar}}           &                   & \dwn{q_{\hbar}}\\
 \widehat{\g}/_{[[\hbar]]}&                   &C^{\infty}(M)[[\hbar]]/_{[[\hbar]]}.
\end{array}$$
\ \\

On the other hand, it is clear that $
\widehat{\g}/_{[[\hbar]]}\sim {\g}\oplus\RR=\widetilde{\g}$ and
$C^{\infty}(M)[[\hbar]]/_{[[\hbar]]}\sim C^{\infty}(M)$.

Thus we can consider that the quantum momentum map $\widehat{\J}$
recovers the $Ad^*$-equivariant classical momentum map
$\widetilde{\J}_0$ at the classical limit.

\begin{remark}
Let us define
$\dot{\J}=\widehat{\J}|_{{\g}\oplus\{0[[\hbar]]\}}\in
C^{1}({\g},C^{\infty}(M))[[\hbar]])$. By definition,
$\dot{\J}(\xi)=\widehat{\J}(\xi,0)={\J}(\xi)$. Then, is clear that
$\dot{\J}$ is not a Lie algebra homomorphism and recovers
${\J}_0$.
\end{remark}

\begin{remark}
It is clear that when anomalous quantum momentum maps $\breve{\J}$
and $\dot{\J}$ there exists they agree.
\end{remark}
\ \\

In order to relate the existence of both we analyze the conditions
established in Proposition (\ref{existencia de momento anomalo})
and Corollary (\ref{existencia de momento quantico extendido}).

According to Proposition (\ref{existencia de momento anomalo}),
there exists $\dot{\J}$ if and only if there exists  ${\J}_+ \in
\hbar \,C^{1}({\g},C^{\infty}(M))[[\hbar]]$ such that $i_{X_{\xi}}
\Omega= d\, {\J}_+(\xi)$. On the other hand, according to
Corollary (\ref{existencia de momento quantico extendido}), there
exists $\breve{\J}$ if and only if there exists ${\bf
\Jtil_{+}}\in\hbar \ C^{1}(\gtil,C^{\infty}(M))[[\hbar]]$ such
that
$$i_{\Xtil_{(\xi,a)}}\Omega=d{\bf \Jtil_{+}}(\xi,a)$$ and
$$\Omega\left(\Xtil_{(\xi,a)},
\Xtil_{(\eta,b)}\right)=\left(\delta_{\rhotil_{c}}{\bf
\Jtil_{+}}\right)_{((\xi,a),(\eta,b))}.$$

We can see that the existences of ${\J}_+\in \hbar
\,C^{1}({\g},C^{\infty}(M))[[\hbar]]$ and  ${\bf
\Jtil_{+}}\in\hbar \ C^{1}(\gtil,C^{\infty}(M))[[\hbar]]$ are
related in the following way.

\begin{lemma}
There exists ${\bf \Jtil_{+}}\in\hbar \
C^{1}(\gtil,C^{\infty}(M))[[\hbar]]$ such that
$i_{\Xtil_{(\xi,a)}}\Omega=d{\bf \Jtil_{+}}(\xi,a)$ if and only if
there exists ${\bf \J_{+}}\in\hbar \
C^{1}({\g},C^{\infty}(M))[[\hbar]]$ such that
$i_{X_{\xi}}\Omega=d{\bf \J_{+}}(\xi)$.
\end{lemma}

{\bf Proof.}

If there exists $\widetilde{\J}_+$ that satisfies
$i_{\Xtil_{(\xi,a)}}\Omega = d\widetilde{\J}_{+}(\xi,a)$, it is
clear that ${\bf J}_+ : {\g} \rightarrow C^{\infty}(M)[[\hbar]]$
defined as ${\J}_+(\xi)=\widetilde{\J}_+(\xi,0)$ satisfies that
$i_{\Xtil_{\xi}}\Omega = d{\bf \Jtil_{+}}(\xi)$.

Reciprocally, if there exists ${\bf J}_+ \in
C^{\infty}(M)[[\hbar]]$ such that verifies $i_{\Xtil_{\xi}}\Omega
= d{\bf \Jtil_{+}}(\xi)$, it is clear that ${\widetilde{\J}_+} :
\widetilde{\g} \rightarrow C^{\infty}(M)[[\hbar]]$ defined as
$\widetilde{\J}_{+}(\xi,a)={\bf J}_+(\xi)$ satisfies
$i_{\Xtil_{(\xi,a)}}\Omega = d{\bf \Jtil_{+}}(\xi,a)$.\
$\triangle$
\ \\

\begin{remark}
Let us notice that the existence of $\breve{\J}$ is more strong
that the existence of $\dot{\J}$.
\end{remark}

\end{document}